\documentstyle[aps,preprint]{revtex}
\def\CA{{\cal A}}	\def\CB{{\cal B}}

\newcommand\Z{{\rm Z\!\!Z}}

\newcommand{\R}{{\rm I\!R}}

\def\ee{\hbox{e}}
\begin{document}

\title{Dbrane Phase Transitions and Monodromy in $K$-theory}
\author{Sergei Gukov\thanks{On leave from the Institute
of Theoretical and Experimental Physics and the L.~D.~Landau
Institute for Theoretical Physics.} and Vipul Periwal}
\address{Department of Physics, Princeton University,
Princeton, New Jersey 08544}
\date{PUPT-1884; ITEP-TH-39/99; hep-th/9908166}
\maketitle
\tightenlines
\begin{abstract}
Majumder and Sen have given an explicit construction of a first order
phase transition in a non--supersymmetric system of Dbranes that occurs
when the $B$ field is varied.
We show that the description of this transition in terms of $K$-theory
involves a bundle of $K$ groups of non--commutative algebras over
the K\"ahler cone with nontrivial monodromy.
Thus the study of monodromy in $K$ groups associated with quantized
algebras can be used to predict the phase structure of systems of
(non--supersymmetric) Dbranes.
\end{abstract}
\bigskip


A major development in string theory over the past couple of years has been
an increasing understanding of non--supersymmetric Dbranes and their
dynamics, initiated largely by Sen \cite{many}.  The relevance of
$K$-theory for the classification of Dbrane charges was pointed out by
Witten \cite{wk}\ following earlier work of Minasian and Moore \cite{mm}.
Reviews of these developments are \cite{sen,lerda,szabo}.

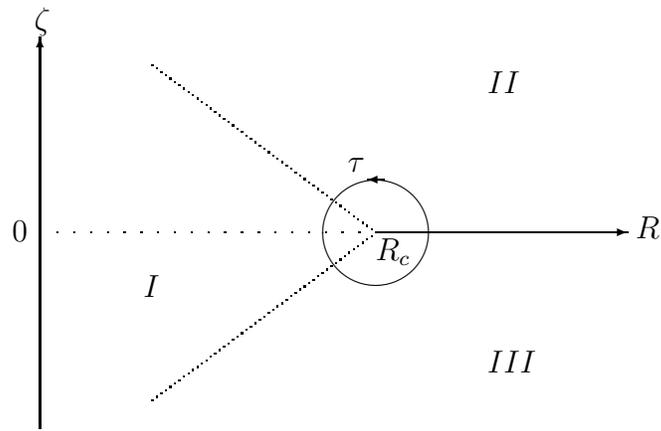
\begin{figure}
\setlength{\unitlength}{0.9em}
\begin{center}
\begin{picture}(22,15)
\put(0,0){\vector(0,1){14}}\put(-0.2,14.3){$\zeta$}
\qbezier[20](0,7)(6,7)(12,7)
\put(12,7){\vector(1,0){9}}\put(21.3,6.8){$R$}\put(12,6){$R_c$}
\qbezier[50](12,7)(8,10)(4,13)
\qbezier[50](12,7)(8,4)(4,1)
\put(12,7){\circle{4}}
\put(12.3,8.9){\vector(-1,0){0.6}}
\put(11,9.2){$\tau$}
\put(-1,6.7){$0$}\
\put(3.7,4.7){$I$}
\put(16,12){$II$}
\put(16,2){$III$}
\end{picture}\end{center}
\caption{Phase diagram in the $R$ -- $\zeta$ plane.}
\label{fig}
\end{figure}

The motivation for the present work was an intriguing observation
made by Majumder and Sen \cite{senmaj}.
Using an exact conformal field theory description\cite{sencft}\
of non-supersymmetric Dbranes
on a K3 surface,
they found three different phases in the region on the moduli space
where the surface looks like an orbifold $X=T^4/\Z_2.$
The phase diagram in Fig. \ref{fig} represents different states
(labeled as $I$, $II$ and $III$) of two D-branes
\footnote{ The spectrum of D$p$-branes in Type II string
theory is periodic under $p \to p+2$, so the value of $p$ will
not be important in our discussion, as long as $p \ge 2$.
In $K$-theory this periodicity is just  Bott periodicity.}
wrapped over exceptional 2-cycles $S_1$ and $S_2$ in $X,$
which correspond to fixed points in the orbifold limit.
In   phase $I$ the two Dbranes recombine into a single Dbrane wrapped
over a non--supersymmetric cycle
with the homology class $[S] = [S_1] + [S_2].$
This happens when the minimal radius of $T^4$ becomes less than
the difference of the $B$-field fluxes through $S_1$ and $S_2$,
$\zeta < R$ with the appropriate normalization.
{}From the analysis of the tachyon potential\cite{senmaj}
\begin{equation}
	V(\alpha) \propto \left({1\over 4}(R_{c}-R)\cos(\alpha\pi) +
	\zeta\cos({1\over 2}\alpha\pi)\right)
\end{equation}
($\alpha$ is the parameter labelling the marginal deformation,
normalized so that $\alpha=0$ mod 2 represents the pair of Dbranes
and $\alpha=1$ mod 2 represents the non--BPS brane) one can see that
transitions between regions $I$ and $II$ and between regions
$I$ and $III$ are second order. There is no discontinuous jump
in the location of the minimum of the tachyon potential in these
transitions.
On the other hand, as we pass through the phase boundary between
regions $II$ and $III$ the Dbranes flip their orientation via a first
order phase transition.
The three phases coexist at the critical radius $R_c$ ($\zeta =0$)
where the anti--periodic tachyonic mode becomes exactly marginal.

We want to understand this phase diagram from the point of view
of $K$-theory and see if such transitions are possible at all.
$K$ groups classify Dbrane charges\cite{mm,wk}.
Therefore, this phase diagram indicates
that a certain element of a $K$ group is mapped to an appropriate inverse
as one crosses the transition line between phases $II$ and $III$.
In other words, it means that a $K$-group element undergoes a monodromy
as one goes around the point $R_c$ in the $R$ -- $\zeta$ plane.
Indeed, because there is no discontinuity along the lines of the second
order phase transitions between phases $II$ and $I,$
and between phases $I$ and $III$, we do not expect
to see these lines in the deformation of the $K$ groups.
What this means precisely, and what can be predicted from
this point of view, constitute the topics of the present paper.

As we explain below, the observation of Majumder and Sen is a
rather general phenomenon in $K$-theory, and we will formulate
the explicit  condition when it takes place for a general space $X.$
It will turn out that the appropriate setting for the analysis
is algebraic $K$-theory  (for introductory treatments see \cite{book,black}),
{\it i.e.} instead of a space $X$ we consider the ring
(algebra) $\CA$ of continuous functions on $X.$
Although the algebraic $K$-theory of $\CA$ is isomorphic  to topological
$K$-theory of $X$ for smooth $X,$ the former has two major benefits:
\begin{itemize}
\item Even if the geometry of $X$ is   singular, $\CA$
may remain a well-defined algebra. For example, if $X$ is
a quotient space with possible fixed point singularities,
$\CA$ is a crossed product algebra \cite{othernoncomm}.
\item Majumder and Sen consider a `blow-up' of $X$ corresponding
to a $B$-field flux through 2-cycles in $X$. Such non-geometric
`defomations' of $X$ have a natural interpretation in terms of
non-commutative deformations of $\CA$ \cite{cds,othernoncomm,seiwit}.
\end{itemize}
An attempt to formulate string theory in algebraic terms can be found
in \cite{vpth}.

We briefly recall the definition of $K$-groups that measure Dbrane charges.
In topological $K$-theory $K^0 (X)$ is defined as the group of
pairs $(E,F)$ of vector bundles modulo the equivalence relation
$(E,F) \sim (E \oplus H, F \oplus H')$ which allows creation and
annihilation of brane--anti-brane pairs with isomorphic gauge
bundles $H$ and $H'.$ In algebraic $K$-theory the Grothendieck group
$K_0 (\CA)$ is defined in a similar way with ``bundles over $X$"
replaced by ``projective modules over $\CA$". In practice, however,
it is convenient to use another (equivalent) definition of
$K_0 (\CA)$ via idempotents in $M_{n}( \CA)$, the set of $n\times n$
matrices, with coefficients in $\CA$. To allow direct sum and tensor
product of idempotents, one actually has to consider the direct limit
$M_{\infty} (\CA) = \lim M_n (\CA)$ with the inductive limit topology.
Then unitary equivalence classes of projection operators $\alpha$:
\begin{equation}
\alpha \ast \alpha = \alpha
\label{projection}
\end{equation}
in $M_{\infty}(\CA)$ form a semigroup $S\equiv S(\CA)$ under addition.
The Grothendieck group associated with this semigroup is constructed
as follows. On $S \times S$ define an equivalence relation
$(a,b) \sim (a',b')$ if $a+b'=a'+b$. Then $K_{0}(\CA) = S\times
S/ \kern-0.3em \sim$.
By definition $K_0$ is a covariant functor, {\it i.e.} any
homomorphism $\phi: \CA \to \CB$ of $C^{\ast}$-algebras $\CA$ and $\CB$
induces a homomorphism $\phi_{\ast}: K_0(\CA) \to K_0 (\CB)$.
If we define a suspension as the set of
continuous functions from the real line to $\CA$,
$\Sigma \CA \equiv C(\R \rightarrow \CA) \cong \CA \otimes C(\R)$,
we may introduce the higher K-group $K_1(\CA) \cong K_0 (\Sigma \CA)$.
Equivalently, one can define $K_{1}(\CA)$ as the abelian group of
equivalence  classes of invertible elements in $M_{\infty}(\CA)$,
with equivalence defined by right translation by elements of
the group that are in the identity component. Complex Bott
periodicity is the statement that $K_{0}(\Sigma^{2} \CA) = K_{0}( \CA)$.

According to Connes, Douglas and Schwarz \cite{cds}, the presence
of a uniform $B$-field can be interpreted in terms of a quantization of
the function algebra of the manifold, \`a la Fedosov \cite{fed}.
Their deep insight has led to numerous developments
\cite{othernoncomm,seiwit} on this theme.
We are interested in a $B$-field which is
not uniform \cite{senmaj}. Since no precise relation to
deformation quantization has been established for this case,
we assume that a general $B$-field leads to the deformed algebra
$\CA$ with the $\ast$ product defined by Kontsevich \cite{kon}.
In fact, as it will become clear in a moment, the results of this
paper do not depend on this assumption; one just has to know
that turning on a $B$-field leads to {\it some} associative
deformation of a product on $\CA:$
\begin{equation}
f\ast_{\tau}g = f g + \tau f \ast_{1}g + \ldots
\end{equation}
where $\tau$ is the deformation parameter.
In the following we will only make the $\ast$ symbol explicit
when we need to emphasize the deformed product.
The $\ast$ product depends not just on $B$ but also on the K\"ahler
form $J.$ However,
this dependence is such that when $B=0 $ the algebra is not expected
to undergo any deformation\cite{cds,seiwit}.
This is perhaps surprising at first sight since the natural variable in string
theory is $B + iJ$, but
it follows from the modular--like invariance of the $\ast$ product
under $T$ duality.

Thus, over each point in the $(R,\zeta)$ plane we have an algebra
in which the product depends on the value of $z \equiv (R,\zeta)$.
We shall take the undeformed algebra in our model to be the
algebra at $z=R_{c}$. Therefore
we define the deformation parameter $\tau \equiv z-R_{c}$.
Computing the $K$ groups of these algebras at each point in
the $z$-plane, we obtain a bundle of $K$ groups.
These $K$ groups at different points on the $(R,\zeta)$ plane
{\it a priori} have little to do with each other.

However, the important point is that $K$ groups are, generally
speaking, {\it rigid} under deformation quantization \cite{rosen}.
This fact is the conceptual reason why BPS states survive turning
on a $B$-field and why BPS conditions in the presence of a $B$-field
are very simple to state in terms of the $\ast$ product \cite{seiwit}.
Now it is easy to see the deformation of (\ref{projection}) under
quantization. Given a projection $p$ in $\CA_{0},$  where
$\CA_{0}$ is the algebra before quantization, we need to find
a projection $p_{\tau}$ in $\CA_{\tau}$ such that
$p_{\tau} \ast_{\tau} p_{\tau}= p_{\tau}$ and
$\lim_{\tau \rightarrow 0} p_{\tau} = p.$
This can be solved recursively as a power series,
or by using the differential equation
\begin{equation}
p'_{\tau} \ast_{\tau} p_{\tau} + p_{\tau} \ast_{\tau} p'_{\tau} +
p_{\tau} \Delta \ast_{\tau} p_{\tau} = p'_{\tau},
\end{equation}
where $\Delta \ast_{\tau}$ denotes the derivative of the $\tau$
dependence in the $\ast$ product, with the obvious boundary
condition at $\tau=0.$ Thus there is a natural manner in which
the fibres of the $K$-group  bundle are all isomorphic, so it
actually is a fibre bundle. Given this isomorphism, we can
meaningfully consider the {\it monodromy} of sections of this fibre
bundle as we move around $\tau =0$.

In supersymmetric cases such monodromies are very well
understood from the relation between the Mukai vector \cite{cds}:
\begin{equation}
\vec Q = {\rm ch} (E) e^{ - {[B] \over 2 \pi i}}
\in H^{{\rm even}} (X, \Z)
\label{mukai}
\end{equation}
and the Chern character of a projective module $E.$
In physics this expression appears in the Chern-Simons coupling
of Ramond-Ramond fields.
Let us explain the origin of the monodromy in algebraic
$K$-theory by a simple example of Type IIB compactification
on a non-commutative torus.
An ordinary torus $T^2$ can be represented as a quotient
space $X = S^1 \times \R / \Z$. The corresponding crossed
product algebra $\CA_0 = C(S^1) \times \Z$ is generated by
two unitary operators that commute with each other.
Its non-commutative deformation $\CA_{\tau} = C(S^1) \times_{\tau} \Z$
is defined by `twisting' the multiplication by a
$\tau \in {\rm Aut} (C(S^1))$: $ag=g \tau (a)$ for all
$a \in C(S^1)$ and $g \in \Z$.
The deformed algebra $\CA_{\tau}$ is Morita equivalent
to a non-commutative torus which is believed to describe
a compactification on a torus with a $B$-field flux \cite{cds}.
Because $\CA_{\tau}$ is still a crossed product algebra,
we can use the Pimsner--Voiculescu exact sequence \cite{pv}
to compute its $K$ groups:
\begin{equation}
\dots \longrightarrow K_0 (C(S^1)) \stackrel{{\rm Id} - \tau}{\longrightarrow}
K_0 (C(S^1)) \longrightarrow K_0 (C(S^1) \times_{\tau} \Z) \longrightarrow
K_1 (C(S^1)) \longrightarrow \dots
\label{pvsequence}
\end{equation}
First of all, from this sequence we see
the result, alluded to above, that $K$-theory is rigid under
deformation quantization. This follows from the fact that the map
${\rm Id} - \tau$ is homotopic to zero, so that the Pimsner--Voiculescu
exact sequence reduces to a short exact sequence and splits.
The same statement is also true in  real $K$-theory \cite{cargese}.
Furthermore, from (\ref{pvsequence}) we find
$K_0 (\CA_{\tau}) = \Z + \tau \Z$, so that charges of odd-dimensional
Dbranes undergo a monodromy:
$$
\vec Q \longrightarrow \pmatrix{1 & 1 \cr 0 & 1} \vec Q
$$
as the $B$-field changes by a period.

In the non-supersymmetric situation considered by Majumder and Sen
\cite{senmaj}, we expect monodromy in going around the point
$R_{c}$ in the $z$ plane.
We wish to stress here that in the moduli space of K3 with
a $B$-field this monodromy does {\it not} correspond to any
non-trivial element of the first fundamental group.
We do not have a sufficiently detailed understanding of
the $K$ group bundle in the Majumder-Sen case to compute
this monodromy exactly. However, we shall now explicitly
construct a simple local model showing how such monodromy arises.

If $\alpha$ is the projection corresponding to phase $II$,
in going around a small loop centered at $R_{c}$ we expect
to find a projection associated with the non--BPS Dbrane
with the opposite charge. What is the projection associated
with such an oppositely charged Dbrane?  The sum of a charge
and the opposite charge should be equivalent to a trivial bundle, but
$-\alpha$ is not a projection so we have to work a little harder to
find an  inverse.
Given a projection $\alpha,$ and a projection $\pi_{n}$
such that $\pi_{n}>\alpha,$ we note that \begin{equation}
(\pi_{n}-\alpha)^{2}=\pi_{n} - \pi_{n}\alpha
-\alpha \pi_{n} + \alpha = \pi_{n }-\alpha
\end{equation}
so $\pi_{n}-\alpha$ is also a projection.  For an appropriate choice
of $\pi_{n}$ this is a natural candidate for the projection associated
with the oppositely charged Dbrane.
The physical motivation for this answer is based on
the description of Dbranes as topological defects in
a gauge bundle of higher dimensional branes \cite{wk}.
We say that two Dbranes carry opposite charges if they are
represented by the gauge bundles $E$ and $F$, such that the
`total' bundle $E \oplus F$ is isomorphic to a trivial bundle.
It is easy to see that $\alpha$ and $\pi_n - \alpha$ satisfy
the expected property provided that $\pi_n > \alpha$.

Now the condition for monodromy around $R_{c}$ is
\begin{equation}
\alpha(\tau \ee^{2i\pi}) = U (\pi_{n}-\alpha(\tau)) U^{\dagger}
\label{monodromy}
\end{equation}
where $|\tau|$ is the radius of a small circle around
$R_{c}$ in the $z$ plane, see Fig. \ref{fig}.
Evaluating at $\tau=0$ we find a condition in
the undeformed $K_{0}(\CA_{0})$ group:
\begin{equation}
\alpha_0 = U (\pi_{n}-\alpha_0) U^{\dagger}\ .
\label{zero}
\end{equation}

As a simple example,   the reader may find it helpful to keep in mind
the following two-dimensional model:
\begin{equation}
\alpha_0 = \left(\matrix{1&0\cr0&0}\right), \quad
\pi_{2}= \left(\matrix{1&0\cr0&1}\right), \quad
U =\left(\matrix{0&1\cr1&0}\right).
\label{example}
\end{equation}
where we assume that $U$ does not depend on $z$.
This may be too strong an assumption in the general case,
but suffices for the local model we construct.

It is clear that order by order one can reconstruct the solution
to (\ref{projection}) and (\ref{monodromy}) as a power series in $\tau$:
\begin{equation}
\alpha (z) = \alpha_0 + \tau \alpha_{1} + \tau^{2} \alpha_{2} +
\ldots\quad .
\end{equation}
The exact form of the solution will depend on the details of
the $\ast$ product. To see this, we write down the first order
equations that follow from (\ref{projection}) and (\ref{monodromy}):
\begin{equation}
\{ \alpha_{1}, U \} =0 ,\qquad
\alpha_1 - \{\alpha_{1}, \alpha_{0}\} = \alpha_{0} \ast_{1} \alpha_{0}.
\label{first}
\end{equation}
Proceeding in the same fashion, we expect to find solutions to
(\ref{monodromy}) only at specific points on the moduli space
(in our case, on the $z$ plane). These points represent the
endpoints of the phase boundaries where some tachyonic modes
become massless \cite{senmaj}.

To summarize, in our local model, we have shown that when the equation
(\ref{monodromy}) has a solution, there is a section of the $K$ group
bundle with  monodromy appropriate for describing the physics of a
first order phase transition between different stable
configurations of Dbranes.
Thus $K$-theory combined with a knowledge of the deformation of the algebra
product as a function of moduli (in this case the difference in the $B$ field
flux through the 2--cycles) can be used to predict the phase structure
of systems of (non--supersymmetric) Dbranes.

In conclusion we mention that there are other kinds of phase
transitions when a sheaf $E$ associated with a Dbrane becomes
unstable or split. The equation of the phase boundary
follows from the Bogomolov condition:
\begin{equation}
\int_X \Big( 2r c_2 - (r-1) c_1^2 \Big) \wedge J^{(n-2)} =0
\label{bogomolov}
\end{equation}
for a sheaf $E$ of rank $r$ over an $n$-dimensional space $X$
to be strictly stable with respect to a K\"ahler form $J.$

\acknowledgements

We have benefited from discussions with R.~Bezrukavnikov, M.~Kontsevich,
J.~Rosenberg, A.~Schwarz, A.~Sen, E.~Sharpe and E.~Witten.
The work of S.G. was supported in part by grant RFBR
No 98-02-16575 and Russian President's grant No 96-15-96939.
The work of V.P. was supported in part by NSF grant PHY-9802484.

\end{document}